\documentclass[11pt]{article}
\usepackage{graphicx}
\usepackage{amsmath}
\usepackage{amssymb}
\usepackage[left=4cm,top=4cm,right=3cm,bottom=5cm]{geometry}

\newcommand{\be}{\begin{equation}}
\newcommand{\ee}{\end{equation}}
\newcommand{\bfig}{\begin{figure}[!ht]\begin{center}}
\newcommand{\efig}{\end{center}\end{figure}}
\newcommand{\ba}{\begin{eqnarray}}
\newcommand{\ea}{\end{eqnarray}}

\begin{document}
\title{Effects of Lateral Diffusion on the Dynamics of  Desorption}
\author{Tjipto Juwono, Ibrahim Abou Hamad, Per Arne Rikvold\\
Department of Physics, Florida State University\\
Tallahassee, FL 32306-4350, USA}
\date{}
\maketitle
\begin{abstract}
The adsorbate dynamics during simultaneous action of desorption and lateral adsorbate diffusion is studied 
in a simple lattice-gas model by kinetic Monte Carlo simulations. It is found that the action of the coverage-conserving 
diffusion process during the course of the desorption has two distinct, competing effects: a general acceleration 
of the desorption process, and a coarsening of the adsorbate configuration through Ostwald ripening. The balance between 
these two effects is governed by the structure of the adsorbate layer at the beginning of the desorption process.\\ 
\\
{\bf Keywords:} Desorption, Diffusion, Coarsening,  Kinetic Monte Carlo simulation
\end{abstract}

\section{Introduction}
\noindent 
Adsorption and desorption of submonolayer structures are nonequilibrium processes of significant scientific and 
technological importance. Examples  can be observed, for instance in systems that undergo 
underpotential deposition \cite{BRO99}. Here we consider the influence of lateral adsorbate 
diffusion on the time evolution of the adsorbate configuration during desorption of a submonolayer from a 
single-crystal surface. While the adsorbate coverage is obviously reduced by the desorption process, it is conserved by the 
process of lateral diffusion. Thus, whereas desorption causes shrinkage of adsorbate clusters of all sizes, lateral 
diffusion may lead to a {\em coarsening\/} of the adsorbate configuration through 
Ostwald ripening \cite{LIF61},\cite{BAR96}. The interplay of these two processes makes the resulting adsorbate 
dynamics quite complex. A likely experimental realization was observed by He and Borguet 
\cite{HEY01} during gold-cluster dissolution on a Au(111) electrode. 
In these experiments, gold atoms were released during a short positive potential pulse onto the substrate,
where they quickly nucleated and formed islands. After the end of the pulse,  small islands tended to decay 
quickly, while large islands initially continued to grow before they also eventually decayed. 

In the present paper we study such a simultaneous process of desorption and lateral diffusion in a simple 
lattice-gas model by kinetic Monte Carlo simulations. The rest of the paper is organized as follows. 
The lattice-gas model and the kinetic Monte Carlo algorithm are introduced in Sec.~\ref{sec-2}, 
the different initial configurations used for the desorption process are discussed in Sec.~\ref{sec-3}, 
our numerical results on desorption without and with lateral diffusion are 
presented in Sec.~\ref{sec-4}, and a brief discussion and conclusions are given in Sec.~\ref{sec-5}. 

\section{Model and Methods}
\label{sec-2}
\noindent Our aim is to study how the cluster size distribution of the initial
configuration influences the effects of lateral diffusion on the 
dynamics of desorption from a submonolayer of adsorbed atoms or molecules.
The simplest kind of model for submonolayer adsorption, desorption, and diffusion 
is a lattice-gas model with only nearest-neighbor interactions.
The grand-canonical effective Hamiltonian defining this model is
\be
\mathcal{H}=-\phi\sum_{\langle i,j \rangle}c_i(t)c_j(t)-\mu\sum_i c_i(t) \;,  
\label{hamiltonian}
\ee
\noindent where the first sum runs over all nearest-neighbor pairs of lattice sites, 
and $\phi > 0$ denotes an attractive nearest-neighbor interaction suitable for describing underpotential deposition.
Here, $c_i$ is the occupation variable at site $i$, with $c_i=0$ if empty and $c_i=1$ if occupied.
The second sum runs over all sites, and $\mu$ is the electrochemical potential.  
The value $\mu_0 = -2 \phi$ for a square lattice corresponds to equilibrium coexistence between a low-coverage and a 
high-coverage phase at a discontinuous phase transition. 
This Hamiltonian is equivalent to a nearest-neighbor Ising spin model~\cite{FRA06}. 

The simulations reported here
were performed on a $256 \times 256$ square lattice with periodic boundary conditions  
at a temperature $T=0.8T_c$ with $T_c$ the exact critical temperature of the Ising  lattice-gas model~\cite{ONS44}. 
Energy and temperature units are chosen
such that  Boltzmann's constant $k_B=1$.
The surface coverage (particle concentration or fraction of occupied 
sites on the lattice) is given by $\theta=N_p/N$, where $N_p$ is the
total number of occupied sites and $N$ is the total number of sites on the lattice. 
Each time we apply an elementary step, the system energy changes from $E_{\rm A}$ to $E_{\rm B}$. 
To accomplish this, the system has to overcome a local energy maximum $E_{\rm H}$. We define the height
of the corresponding energy barrier $\Delta$ by the Butler-Volmer type expression \cite{FRA06},
\begin{eqnarray}
\Delta &=& E_{\rm H} - E_{\rm av} \nonumber \\
       &=& E_{\rm H} - \frac{1}{2}(E_{\rm A} + E_{\rm B}) \;.
\label{eqndelt}
\end{eqnarray}
\noindent The energy difference between $E_H$ and $E_A$ is
\begin{eqnarray}
\Delta {\tilde E} &=& E_{\rm H} - E_{\rm A} \nonumber \\
                  &=& \frac{1}{2}(E_{\rm B} - E_{\rm A}) + \Delta \;.
\label{eqndele}
\end{eqnarray}
\noindent We then define the transition rate as
\begin{eqnarray}
R_{\rm A \rightarrow \rm B}= \exp\left(- \left (\Delta/T \right) \right)
\exp\left(-\left( E_{\rm B} -
E_{\rm A} \right)/2T \right) \;.
\label{eqnrats}
\end{eqnarray}

We consider two kinds of elementary steps: (1) an adsorption/desorption
step, and (2) a diffusion step to a nearest-neighbor site. We apply a
rejection-free ($n$-fold way) algorithm~\cite{BOR75,NOV01,GIL72,GIL76},
in which we update the clock after every accepted configuration change.
The energy barrier for an adsorption/desorption step is $\Delta_{\rm ads/des}$,
and for diffusion it is $\Delta_{\rm d}$. The absolute values
of these energy barriers are not important for our simulations (they are
important if we want to relate the simulation time to physical time).
Our simulations are controlled by the ratio between adsorption/desorption
and diffusion steps. This is achieved by fixing $\Delta_{\rm ads/des}$ at
a constant value throughout all simulations and varying  $\Delta_{\rm d}$.
Attachment between adjacent adparticles is instantaneous in this model. 

During the simulation, we sample the time development of the coverage $\theta$. 
At certain coverages we calculate the
cluster-size distribution using the Hoshen-Kopelman algorithm~\cite{HOS76}.
Using this algorithm we calculate the number density $\rho(s)$ which is
the probability of finding on a randomly chosen site the center
of a cluster of  size $s$.
During the simulation we also measure the correlation length,
which for the square-lattice attractive lattice-gas model can be calculated as~\cite{DEB57,JUW12}
\be
\xi(\theta, \Sigma)=\frac{4\theta(1-\theta)}{\Sigma/N} \;, 
\label{debye}
\ee
where $\Sigma$ is the total number of broken bonds at a given coverage $\theta$.

We start from an empty lattice and equilibrate at negative
chemical potential ($\mu - \mu_0 < 0$) to achieve a very low coverage before
switching to $\mu - \mu_0 > 0$ for adsorption until a coverage
cutoff $\theta_{\rm cutoff}$ is reached. The configuration at
$\theta_{\rm cutoff}$ then becomes the initial configuration for the desorption processes with
$\theta_{\rm init}=\theta_{\rm cutoff}$. 
The coverage cutoff is chosen to be $\theta_{\rm init}=0.35$, well below
the random percolation threshold. We prepare four classes of initial
configurations by applying four different chemical potentials  during the adsorption stage:
$\mu-\mu_0=0.4,~1.0,~2.0,~{\rm and}~10.0$.
Figure~\ref{initdrop} shows typical snapshots of the different initial configurations.

For each run of the simulation, we performed adsorption until
$\theta = \theta_{\rm cutoff}$,  immediately followed by desorption until
a low-coverage equilibrium was reestablished. The simulation time $t=0$ is defined
as the beginning of the desorption stage. During the adsorption stage, we fixed the
adsorption/desorption barrier at $\Delta_{\rm ads/des}=15$ and
turned off the diffusion by setting $\Delta_{\rm d}=150$.

As mentioned above, we varied the chemical potential for the adsorption stage
 to obtain different initial configurations for the desorption process.
For the desorption stage we fixed the chemical potential at $\mu-\mu_0=-0.4$
and the adsorption/desorption barrier at
$\Delta_{\rm ads/des}=15$. To study the effects of diffusion, we performed two
different sets of runs for the desorption stage; one without diffusion
($\Delta_{\rm d}=150$) and the other with relatively fast
diffusion ($\Delta_{\rm d}=8$).
During each desorption run we measured $\theta(t)$, $\xi(\theta)$, and $\rho(s)$.
We averaged the results of 100 independent runs for each simulation setup.

The results of the desorption simulations are displayed in the figures, showing
together the results without and with diffusion. We have four sets of results, corresponding to
the four different classes of initial configurations resulting from the different chemical
potentials used during the adsorption stage. By comparing the results for the simulations without and with diffusion,
we deduce how diffusion affects the dynamics of the adsorbate configuration during the desorption process.

\section{Initial Configurations}
\label{sec-3}
\begin{figure}[!ht]
\begin{center}
$$
\begin{array}{cccc}
    \xi=4.96 & \xi=2.93 & \xi=2.15 & \xi=1.55 \\
   {\includegraphics[scale=0.4]{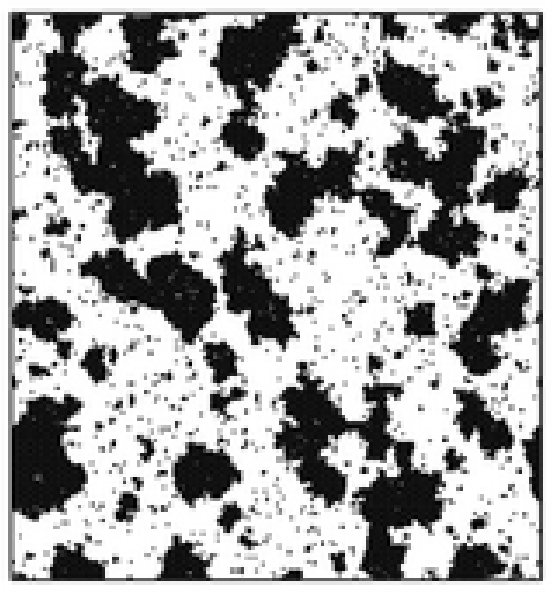}} &
    {\includegraphics[scale=0.4]{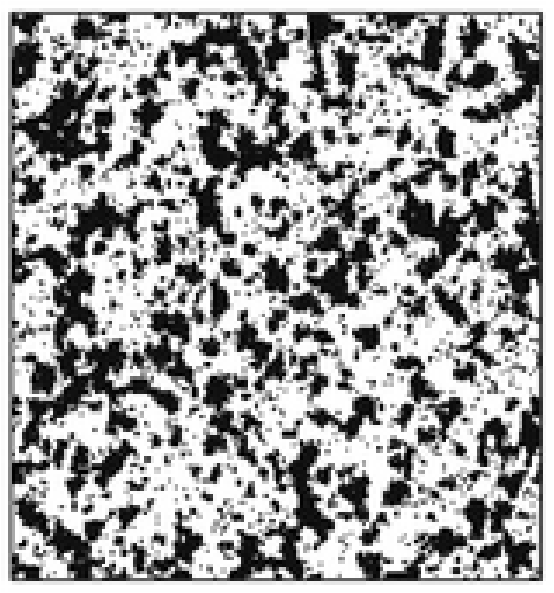}} &
    {\includegraphics[scale=0.4]{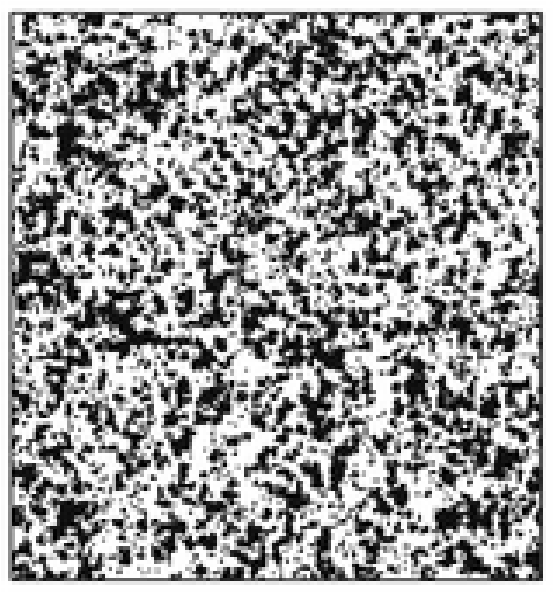}} &
    {\includegraphics[scale=0.4]{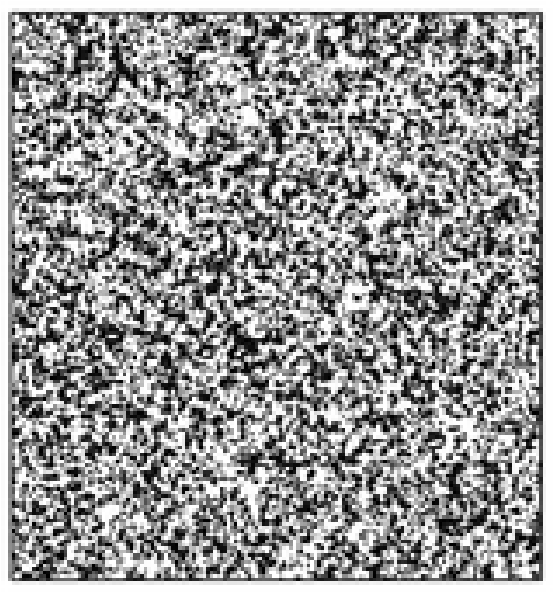}} \\
\end{array}
$$
\end{center}
\caption{Typical initial snapshots prepared by adsorption at different chemical potentials, $\mu - \mu_0 = 0.4$, 1.0, 
2.0, and 10.0 from left to right.}
\label{initdrop}
\end{figure}

\begin{figure}[!ht]
\begin{center}
    {\includegraphics[scale=0.4]{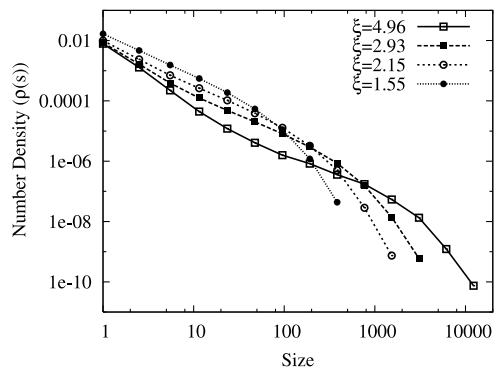}}
\end{center}
\caption{Number density plots $\rho(s)$ 
for initial configurations prepared by adsorption, similar to the ones shown in Fig.~\protect\ref{initdrop}.}
\label{initconfig}
\end{figure}

\noindent Figure~\ref{initdrop} shows typical snapshots of the initial
configurations that we used for the desorption simulations.  
Each of these typical configurations yields a correlation length,
obtained from averaging 100 statistically similar initial configurations.
Each simulation setup is identified by this initial correlation length.
Figure~\ref{initdrop} 
illustrates the fact that within the constraint of our simulation
setup, larger correlation length at a given coverage results in
a cluster size distribution dominated by larger clusters. 
Figure~\ref{initconfig} shows number density histograms for the initial configurations. 
These figures show that we have four initial configurations with size  distributions that span 
(a) $1 \le s < 12000$ for ${\xi_{\rm init}=4.96}$, 
(b) $1 \le s < 3000$  for ${\xi_{\rm init}=2.93}$, 
(c) $1 \le s < 1500$  for ${\xi_{\rm init}=2.15}$, and
(d) $1 \le s < 400$  for ${\xi_{\rm init}=1.55}$.

\section{Simulation Results}
\label{sec-4}
\noindent As the adsorbate configurations changed during the course of the desorption process,
we measured the coverage without and with diffusion. 
Figure~\ref{coverages} shows $\theta(t)$ produced by desorption starting from each of the initial configurations. 
As expected, a larger initial correlation length results in slower desorption as the
initial configuration is dominated by larger clusters.

To quantify the effect of diffusion on $\theta(t)$, we define 
$\Delta t_{\rm 1/2}=t_{\rm 1/2}-t_{\rm 1/2}'$. 
Here the half-time $t_{\rm 1/2}$ is defined as the time for the system to
reach a coverage of $\theta_{\rm init}/2=0.175$ without diffusion, and $t_{\rm 1/2}'$ 
is the half-time with diffusion. We used $\Delta t_{\rm 1/2}$ to measure
the acceleration effect due to the diffusion. Negative $\Delta t_{\rm 1/2}$, therefore,
indicates retardation of the desorption process.

To measure the effect of diffusion on the cluster size distribution $\rho (s)$,
we compared the changes without and with diffusion at a given $\theta$ by measuring $\xi(\theta)$ and $\rho(s)$.
Figure~\ref{corlength} shows $\xi(\theta)$ for each of the four initial configurations.
To quantify the effect of diffusion on $\xi(\theta)$, we define
$\left < \delta \right > = \frac{1}{N_d}\sum_{\theta} \left ( \xi'(\theta)
- \xi(\theta) \right )$, where $\xi(\theta)$ and $\xi'(\theta)$ are the correlation lengths at a given
$\theta$ for desorption without and with diffusion, respectively, and $N_d$ is the number of terms in the sum. 
From Fig.~\ref{corlength} it is evident
that diffusion always increases the correlation length $\xi$ at a given $\theta$.
Figure~\ref{diffeffect} shows the results for $\Delta t_{\rm 1/2}$
and $\left < \delta \right >$ for $1.5 < \xi < 8.1$. The inset in Fig.~\ref{diffeffect}(a) 
shows the cross-over to retardation ($\Delta t_{\rm 1/2} < 0$) for small $\xi_{\rm init}$. 

To further explore the changes in the cluster size distribution caused  by diffusion, 
we sampled $\rho(s)$ histograms with and without diffusion 
at $\theta=0.25$, as shown in Fig.~\ref{histograms}. For $\xi=4.96$ 
(fig.~\ref{histograms}(a)), there are no significant changes by diffusion,  indicating
that diffusion in this case only accelerates the desorption, without having any significant 
effect on the size distribution. In Fig.~\ref{histograms}(b,c,d) we see depletion
of small clusters and enhancement of large clusters, indicating coarsening due to diffusion.

\section{Discussion}
\label{sec-5}
\begin{figure}[!ht]
\begin{center}
$$
\begin{array}{cc}
    {\rm~~(A)~\xi_{\rm init}=4.96} & {\rm~~(B)~\xi_{\rm init}=2.93}\\
    {\includegraphics[scale=0.4]{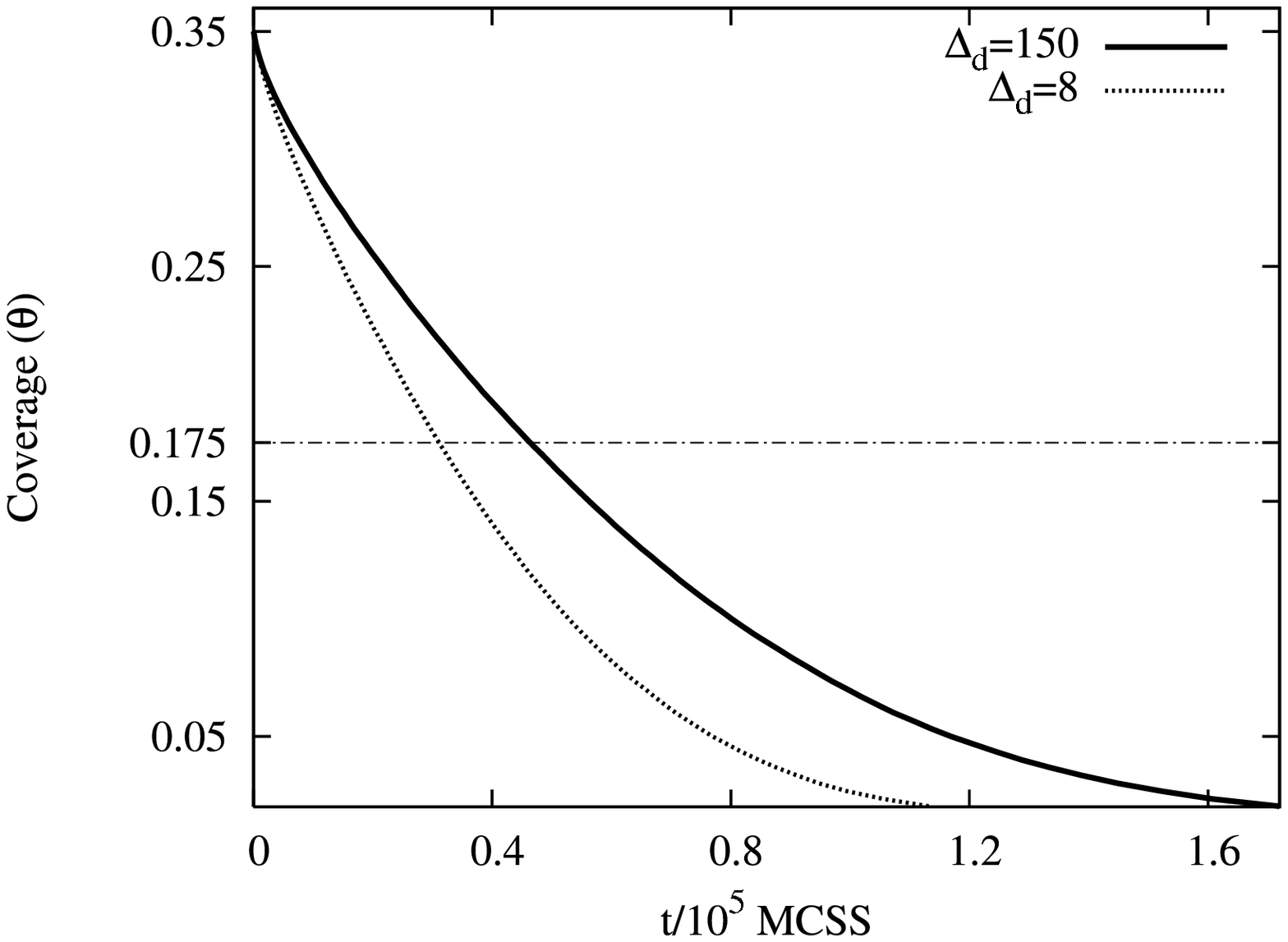}} &
    {\includegraphics[scale=0.4]{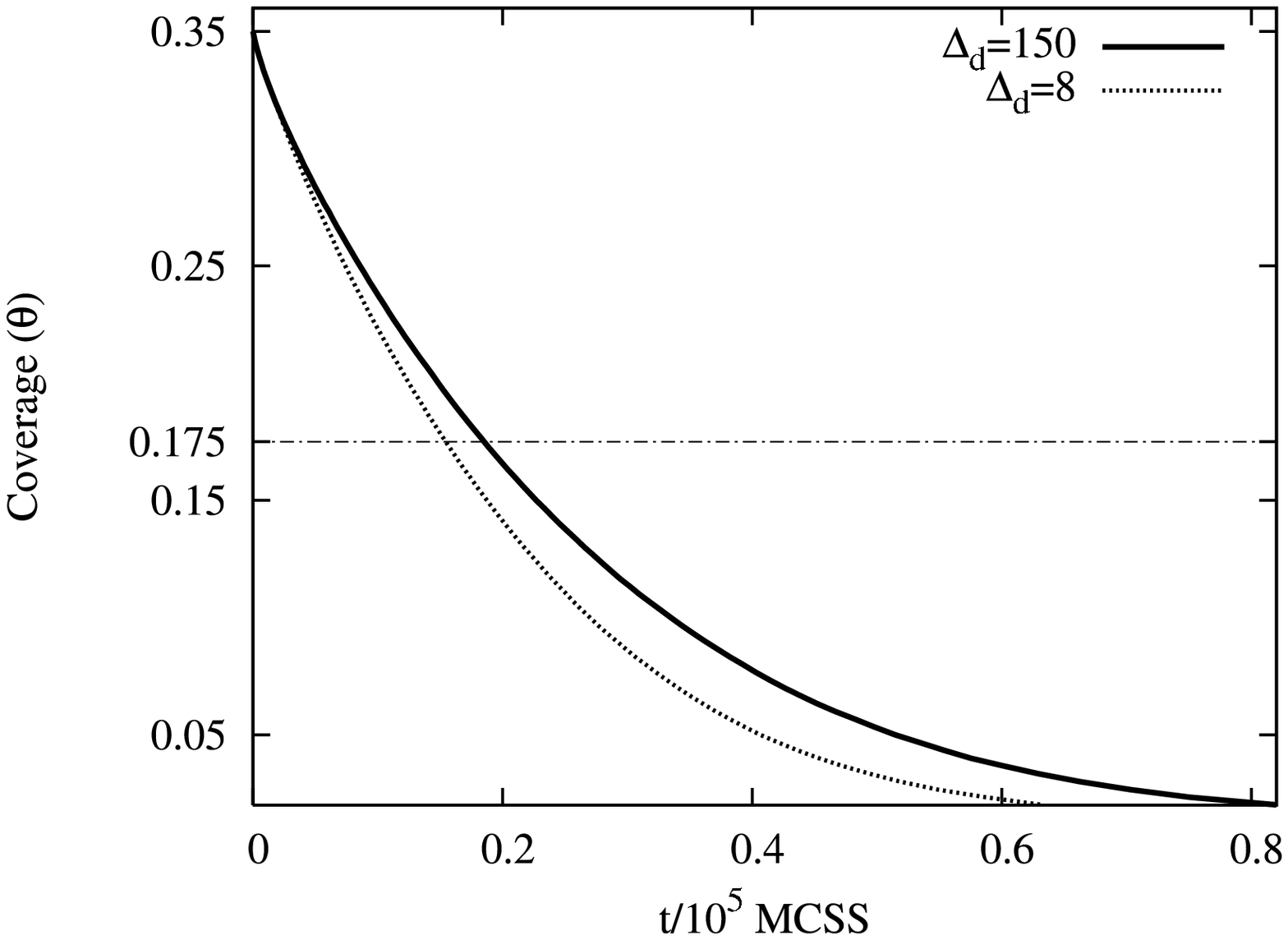}} \\
    {\rm~~(C)~\xi_{\rm init}=2.15} & {\rm~~(D)~\xi_{\rm init}=1.55}\\
    {\includegraphics[scale=0.4]{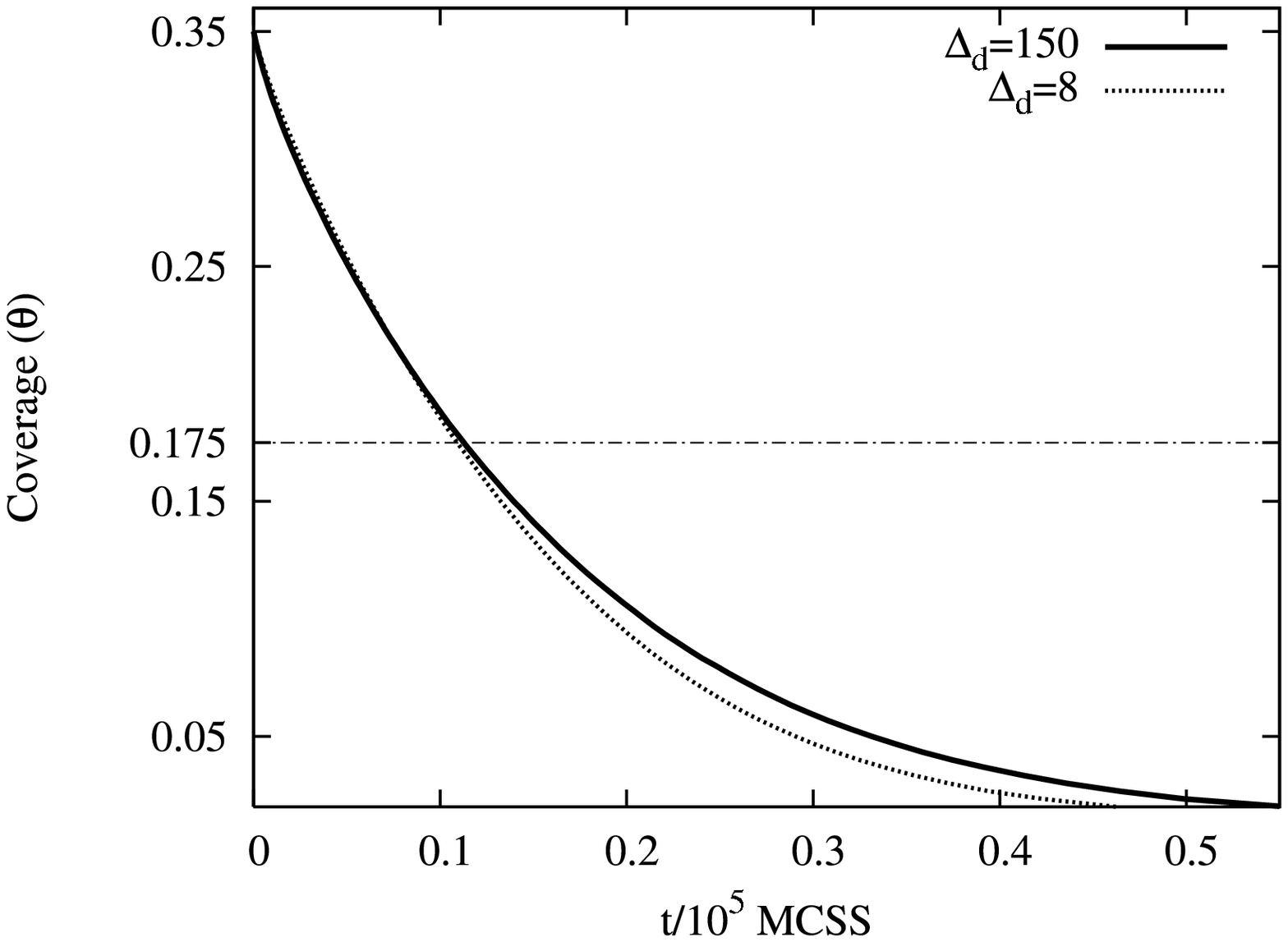}} &
    {\includegraphics[scale=0.4]{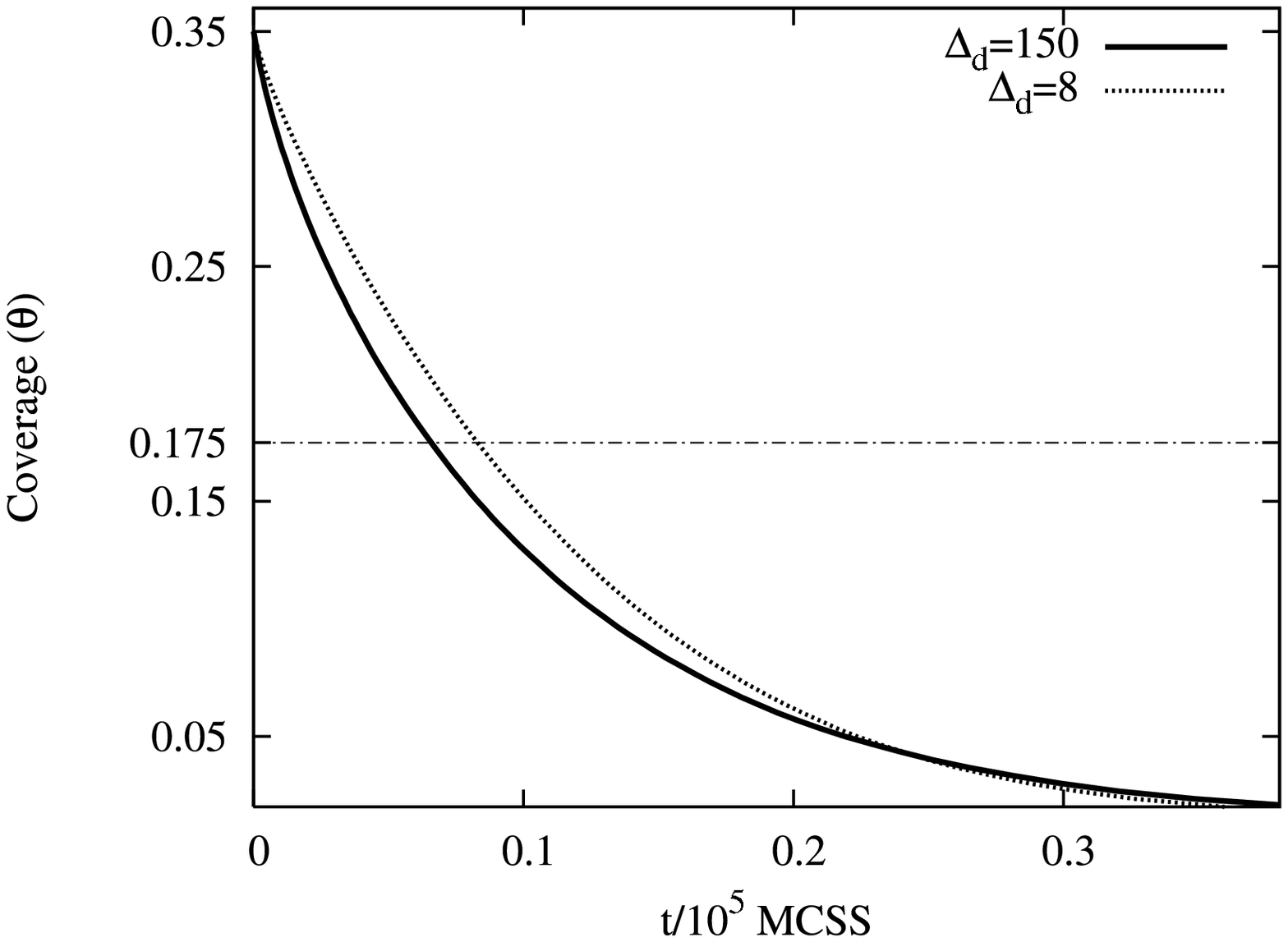}} \\
\end{array}
$$
\end{center}
\caption{\rm The effects of increased diffusion on $\theta(t)$, starting with $\theta_{\rm init}=0.35$ 
and different $\xi_{\rm init}$. Note the different time scales.}
\label{coverages}
\end{figure}

\begin{figure}[!ht]
\begin{center}
    {\includegraphics[scale=0.6]{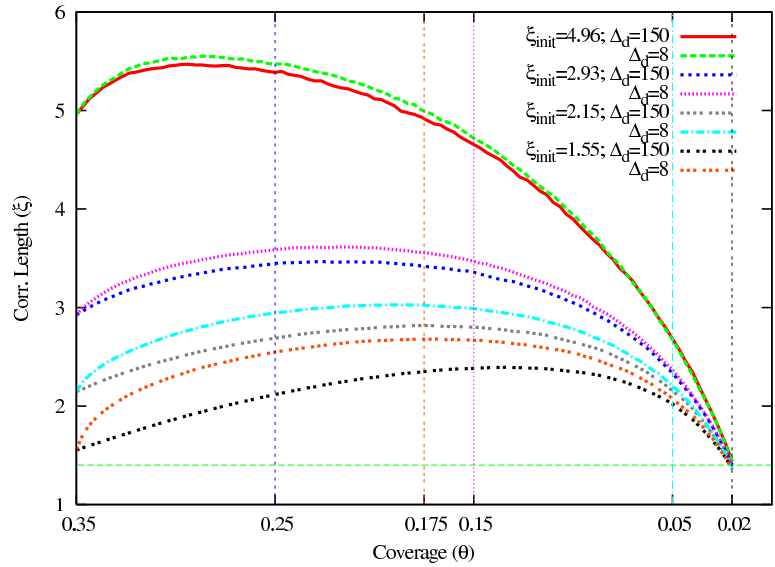}}
\end{center}
\caption{(Color online) Correlation length $\xi$ as a function of
coverage $\theta$, measured without and with diffusion. This plot shows
that diffusion always has a coarsening effect of varying magnitude,
depending on the initial correlation length $\xi_{\rm init}$.
Note that $\theta$ {\em decreases\/} from left to right in this figure, as the time increases.}
\label{corlength}
\end{figure}

\begin{figure}[!ht]
\begin{center}
$$
\begin{array}{cc}
    {\rm (a) } & {\rm (b)} \\
    {\includegraphics[scale=0.4]{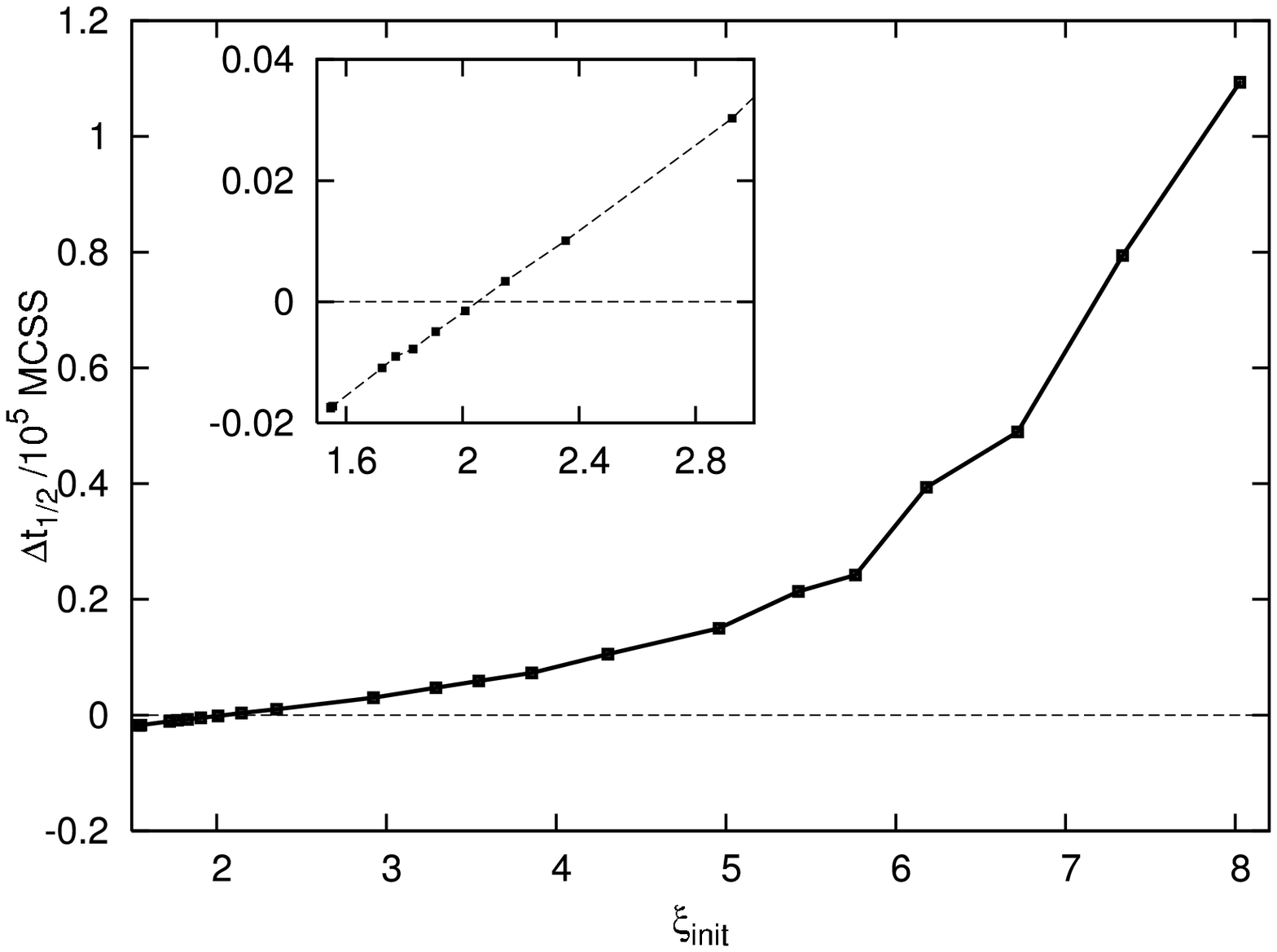}} &
    {\includegraphics[scale=0.4]{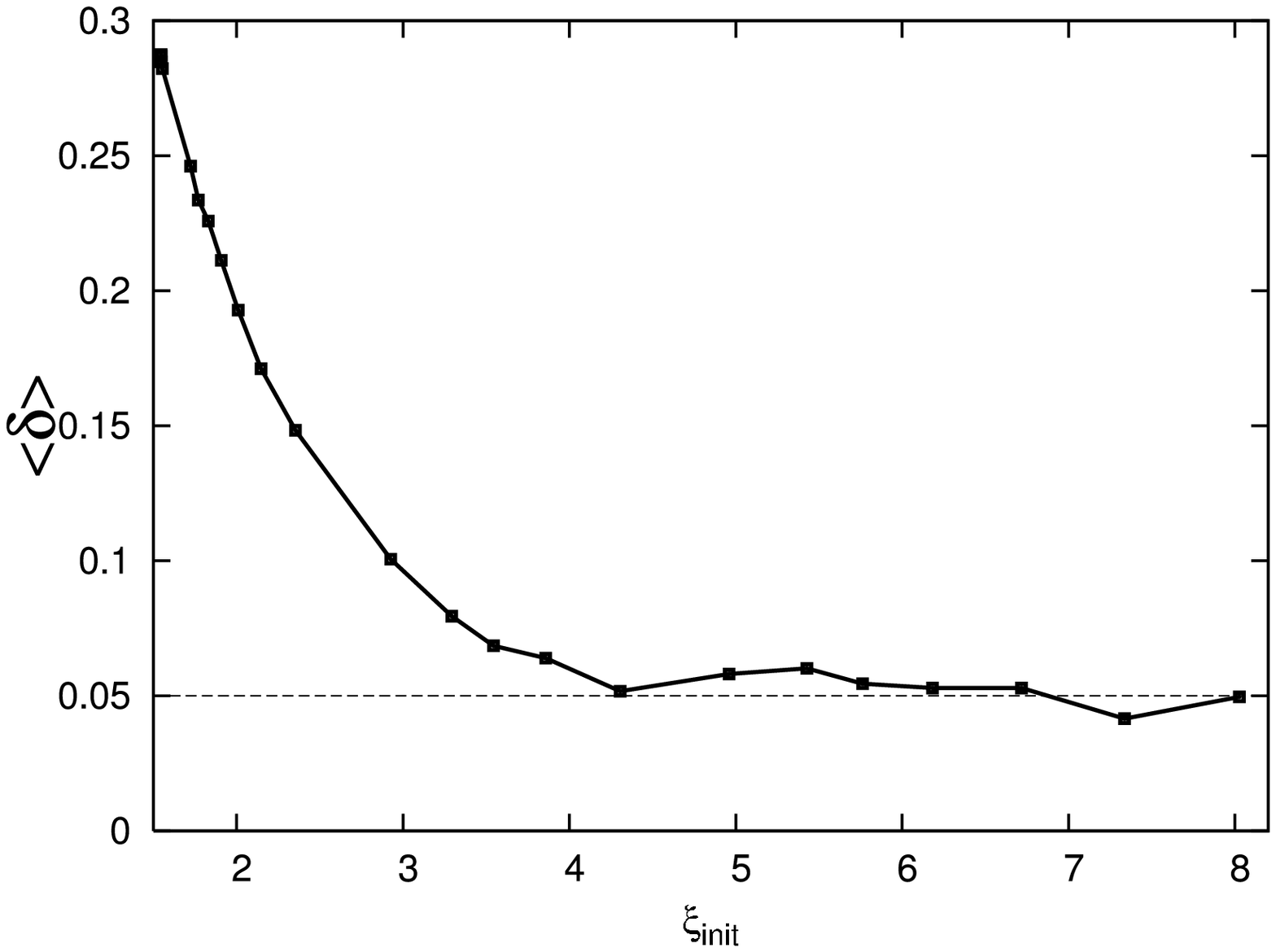}} \\
\end{array}
$$
\end{center}
\caption{Diffusion effects on $\theta(t)$ and $\xi(\theta)$, obtained by measuring 
(a)  $\Delta t_{\rm 1/2}$ and  (b) $\left < \delta \right >$. (See definitions in the text.) Inset in (a): crossover to
 negative  $\Delta t_{\rm 1/2}$.}
\label{diffeffect}
\end{figure}

\noindent From the results for $\theta(t)$ (Fig.~\ref{coverages}) we find
that when starting with larger $\xi_{\rm init}$, the acceleration effect of diffusion
(as shown by $\Delta t_{\rm 1/2}$) increases. This effect decreases when 
we start with smaller $\xi_{\rm init}$. At a certain point, the effect 
crosses over to slowing down the desorption ($\Delta t_{\rm 1/2}$ becomes negative 
for the smallest  $\xi_{\rm init}$ (see Fig.\ref{diffeffect}(a)). 
The measurement of $\xi(\theta)$ (Figs.~\ref{corlength},\ref{diffeffect}), 
on the other hand, shows that $\left < \delta \right >$ is always positive,
and the magnitude of the change increases as we start with smaller  $\xi_{\rm init}$  (see Fig.~\ref{diffeffect}(b)).

\begin{figure}[!ht]
\begin{center}
$$
\begin{array}{cc}
    {\rm (a)} & {\rm (b)} \\
    {\includegraphics[scale=0.4]{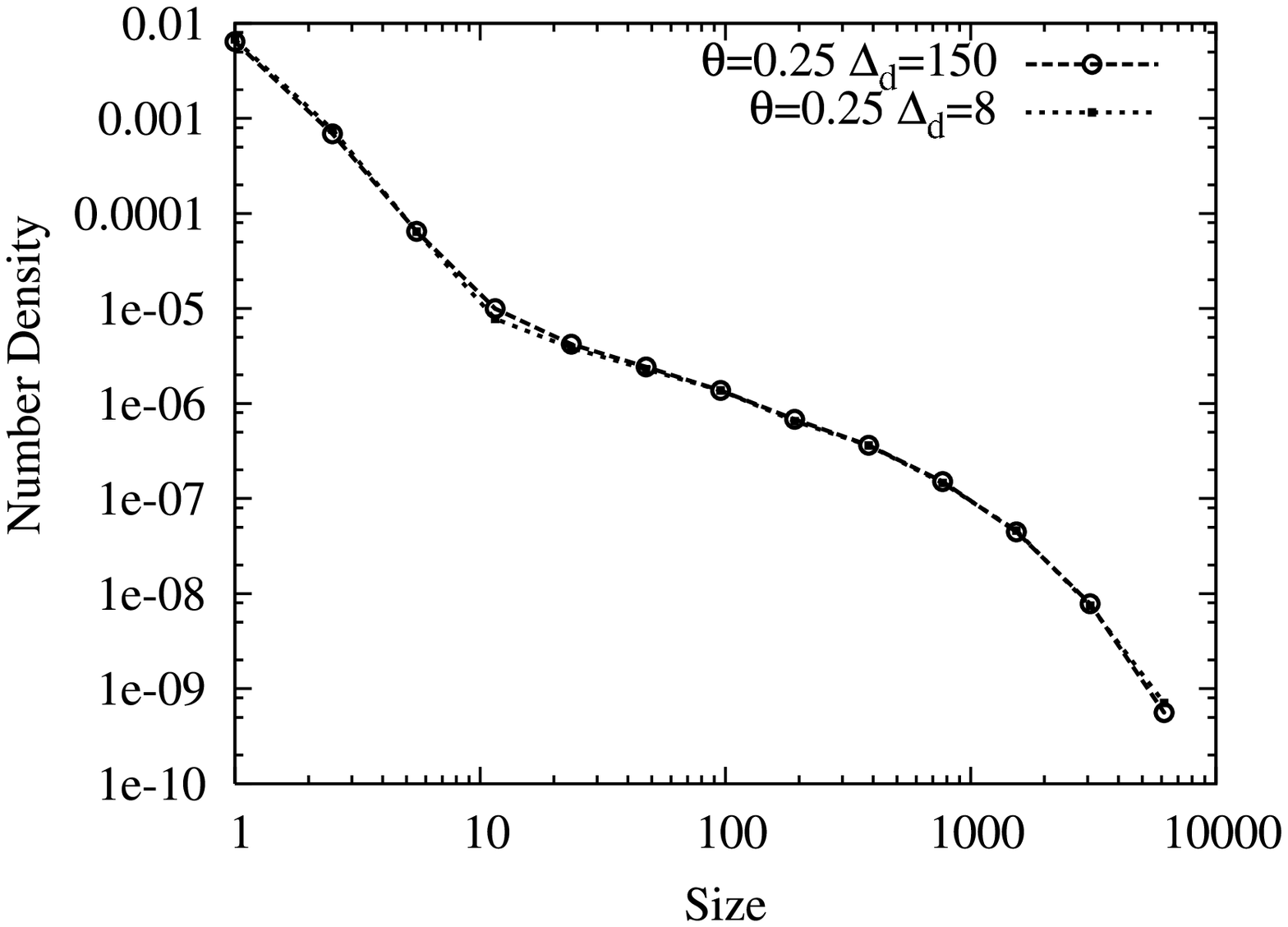}} &
    {\includegraphics[scale=0.4]{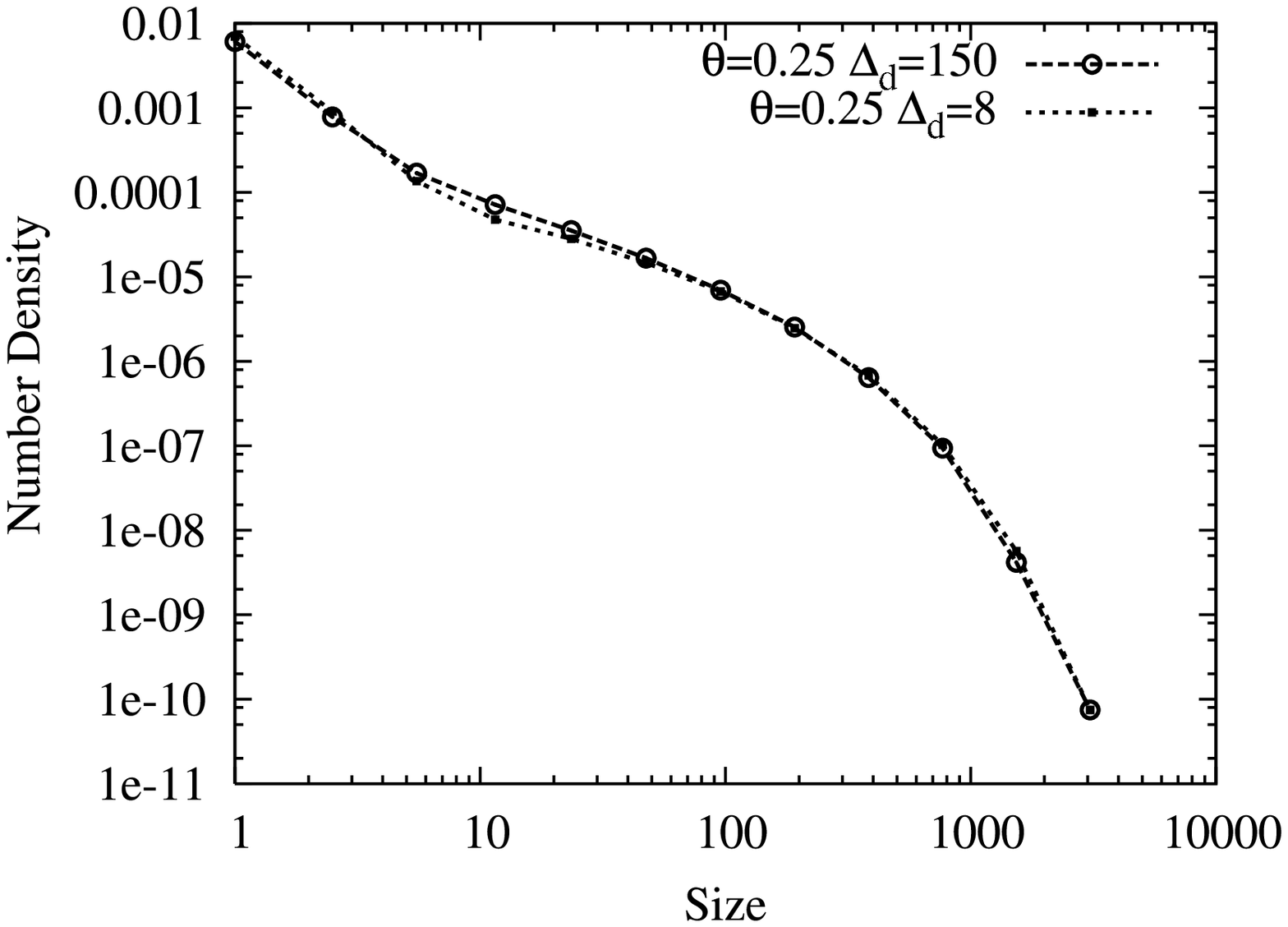}} \\
    {\rm (c)} & {\rm (d)} \\
    {\includegraphics[scale=0.4]{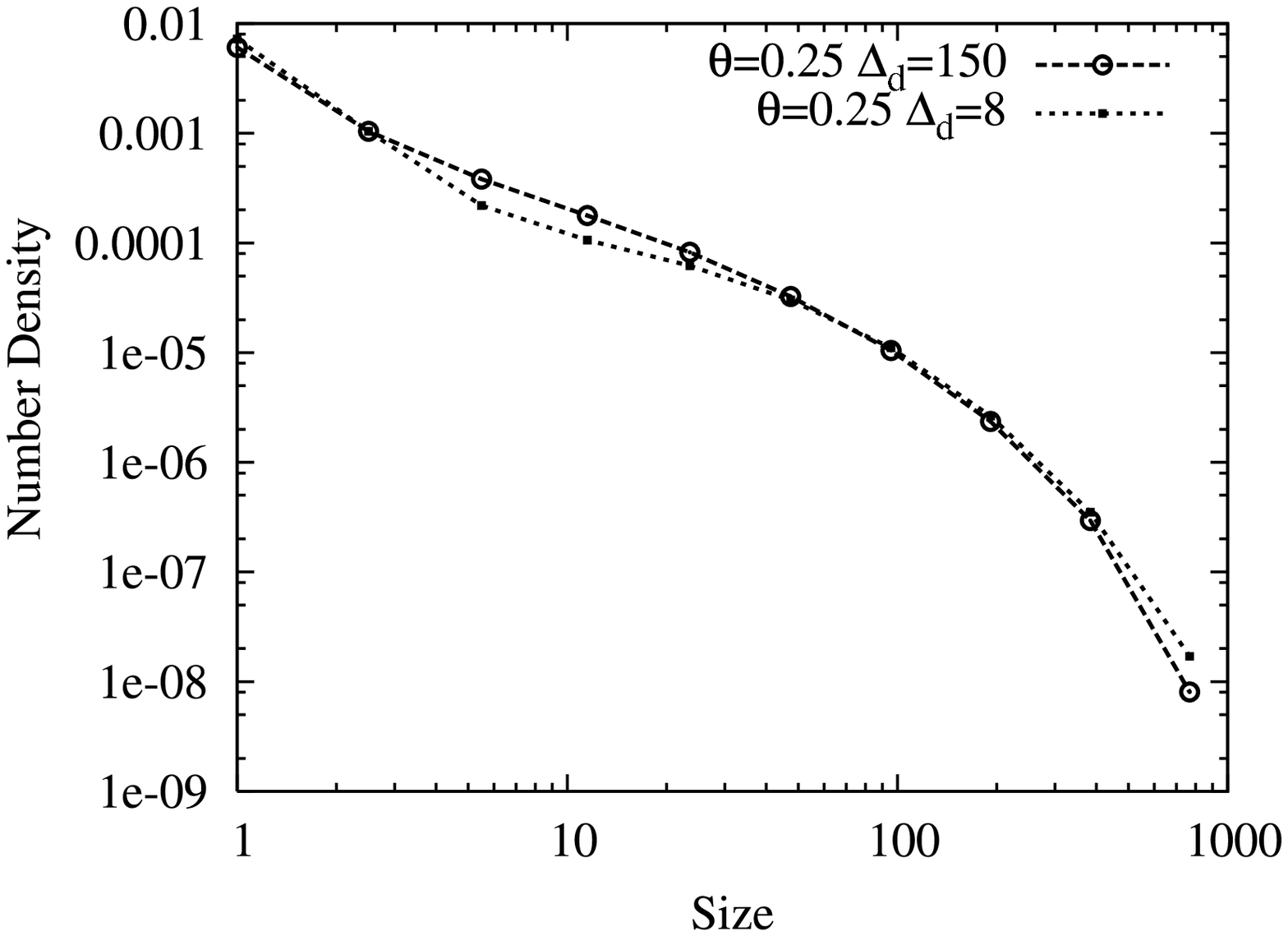}} &
    {\includegraphics[scale=0.4]{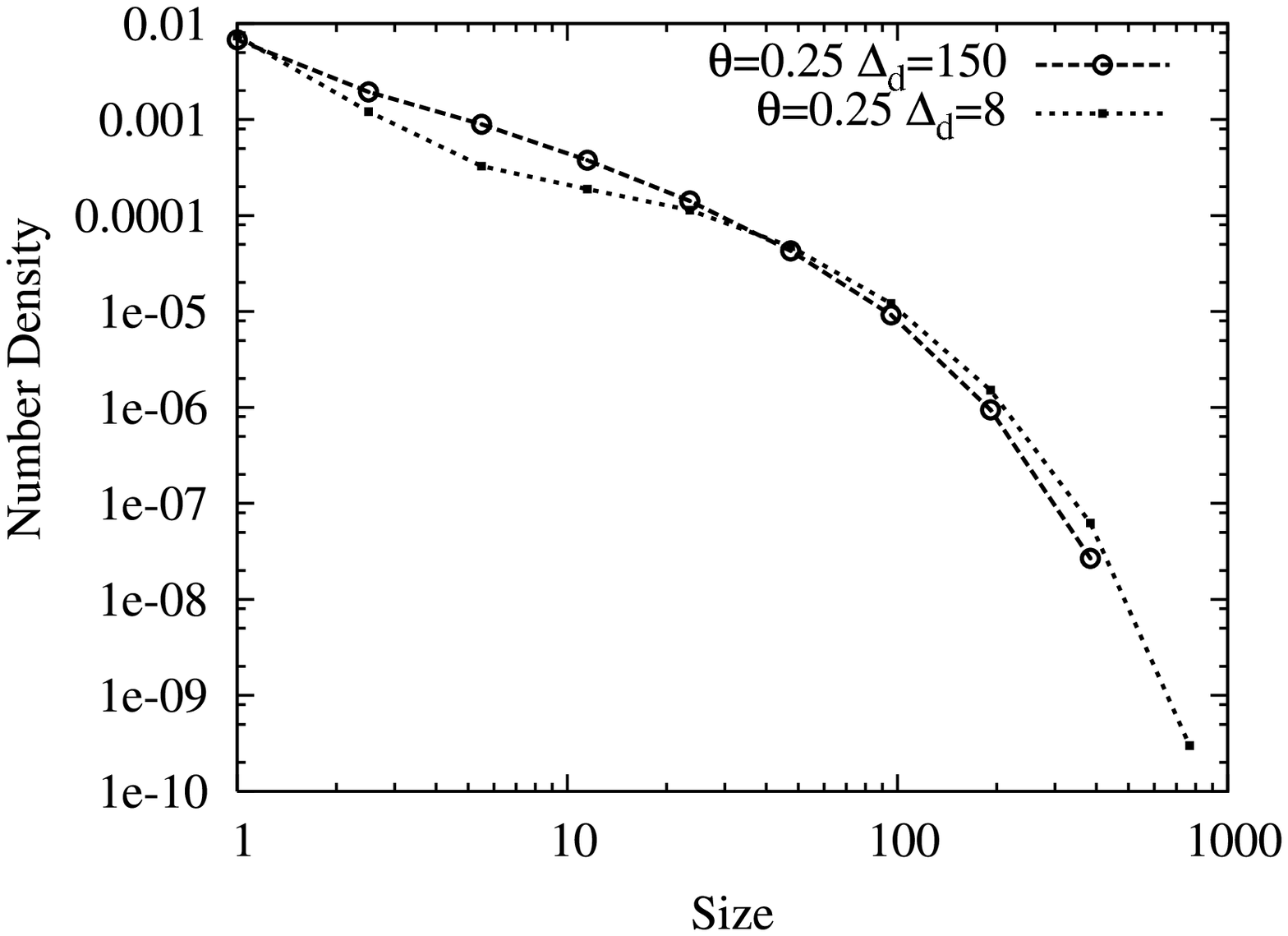}} \\\end{array}
$$
\end{center}
\caption{Size distributions for $\theta=0.25$ without and with diffusion. The simulations
start from $\theta_{\rm init}=0.35$, with 
(a) $\xi_{\rm init}=4.96$, 
(b) $\xi_{\rm init}=2.93$, 
(c) $\xi_{\rm init}=2.15$, and
(d) $\xi_{\rm init}=1.55$.
Note the different scales on both axes in the different parts of this figure.}
\label{histograms}
\end{figure}

Figure~\ref{histograms}(b,c,d) shows depletion of small clusters by diffusion. We associate
this depletion and the positive $\left < \delta \right >$ with a coarsening effect. 
The small positive $\left < \delta \right >$ for the simulation with $\xi_{\rm init}=4.96$ 
indicates a small coarsening effect. In this particular simulation, the acceleration
effect dominates. For simulations with $\xi_{\rm init}=2.93$ and $\xi_{\rm init}=2.15$, we observe
larger  $\left < \delta \right >$ and smaller $\Delta t_{\rm 1/2}$. Finally,
for $\xi_{\rm init}=1.55$, we see large $\left < \delta \right >$ 
with negative $\Delta t_{\rm 1/2}$, which means that the coarsening effect dominates in this case.  

In conclusion, lateral diffusion of adparticles creates two competing effects during desorption: an acceleration
effect and a  coarsening effect. Increasing $\xi_{\rm init}$ decreases the coarsening effect
and increases the acceleration effect, due to enhanced release of particles from the perimeters 
of large clusters. Further details will be presented elsewhere \cite{JUW13}. 

\begin{small}
\section*{Acknowledgements}
\noindent This work was supported in part  by US National Science Foundation grants No.\ 0802288 and 1104829,
and by The State of Florida through the Florida State University Center for Materials Research and Technology (MARTECH). 
\end{small}


\end{document}